\documentclass[%
 reprint,
 amsmath,amssymb,
 aps,
pra,
]{revtex4-1}

\usepackage{color}
\usepackage[table]{xcolor}
\usepackage{graphicx}
\usepackage{dcolumn}
\usepackage{bm}
\usepackage{lipsum}
\usepackage{placeins}
\usepackage{verbatim}

\definecolor{lightlightgrey}{gray}{0.92}

\newenvironment{exer*}
  {\ex}
  {\endex}

\def\be{\begin{equation}}
\def\ee{\end{equation}}
\def\bea{\begin{eqnarray}}
\def\eea{\end{eqnarray}}

\newcommand{\aopd}{\hat{a}^\dagger}
\newcommand{\aop}{\hat{a}}

\newcommand{\bopd}{\hat{b}^\dagger}
\newcommand{\bop}{\hat{b}}

\newcommand{\wt}[1]{\widetilde{#1}}

\newcommand{\phiop}{\hat{\phi}}

\newcommand{\Ephi}{E_C^{(\phi)}}

\newcommand{\Ephitheta}{E_C^{(\phi, \theta)}}

\usepackage{xspace} 

\newcommand{\hairsp}{\hspace{1pt}} 
\newcommand{\ie}{\textit{i.\hairsp{}e.}\xspace} 

\begin{document}

\preprint{APS/123-QED}

\title{Prospects for quantum acoustics with phononic crystal devices}

\author{Patricio Arrangoiz-Arriola}%

\author{Amir H. Safavi-Naeini}
\affiliation{%
Department of Applied Physics, and Ginzton Laboratory, Stanford
University, Stanford, California 94305, USA
}%

\date{\today}

\begin{abstract}

Nanomechanical systems made from phononic crystals can act as highly coherent microwave-frequency circuits at cryogenic temperatures. However, generating sufficient coupling between these devices and microwave superconducting quantum circuits is challenging due to the vastly different length scales of acoustic and electrical excitations. Here we demonstrate a general recipe for calculating these interactions and show that large piezoelectric coupling rates between microwave superconducting circuits containing Josephson junctions and nanoscale phononic resonances are possible, suggesting a route to phononic crystal circuits and systems that are nonlinear at the single-phonon level. 
 
\end{abstract}

\pacs{Valid PACS appear here}

\maketitle

Mechanical filters and resonators, because of their high quality factor and small size compared to electromagnetic components, have been a key part of classical high frequency circuits and systems for the last century~\cite{Campbell1989, Lakin1999}. Advances in microwave-frequency quantum information processing systems have motivated experimental efforts to extend the success of acoustic devices to the quantum realm. In the last decade, a series of experiments have succeeded in coupling superconducting quantum circuits to mechanical resonators~\cite{Irish2003, Cleland2004a, LaHaye2009, OConnell2010, Pirkkalainen2013, Gustafsson2014a, Lecocq2015, Pirkkalainen2015, Aref2016, Magnusson2015}. These approaches have allowed explorations into new regimes of quantum optics~\cite{Aref2016} and enable promising platforms for microwave-to-optical conversion \cite{Safavi-Naeini2011, Bochmann2013a, Balram2015}. However, limiting ourselves to a few figures of merit, there have been no microwave mechanical devices with a convincing advantage over competing electromagnetic systems, where $Q>10^6$ can now be achieved in on-chip resonators~\cite{Megrant2012} and $Q \sim 10^9$ in macroscopic 3D cavities \cite{Reagor2013}. In contrast to room-temperature electrical components, where ohmic heating makes achieving large $Q$'s in small structures impossible, ultracold microwave circuits can leverage the low-loss electromagnetic properties of superconductors, reducing the competitive advantage of mechanical systems. Nonetheless, even in a cryogenic setting, phononic systems can have advantages over electromagnetic quantum circuitry when it comes to coherence time, footprint, and control over crosstalk. Recently, optically-probed microwave frequency mechanical resonators based on phononic crystals have been demonstrated at milliKelvin temperatures with quality factors greater than $10^7$ and footprints on the order of a few square microns~\cite{Meenehan2015}, far exceeding the performance of chip-scale electromagnetic components. The bandgap for acoustic waves in the periodic phononic crystal structure protects localized phonons from tunnelling (clamping loss) and Rayleigh scattering from impurities and defects~\cite{Goryachev2013}. The high level of control over the dispersion of the propagation channels could also enable dense packing of high-$Q$ phononic circuits while limiting crosstalk in large scale systems. Finally, the small mode volume, and the use of high quality single-crystal thin films with low defect densities means that coupling to resonant two-level fluctuators can also be mitigated~\cite{Barends2013, Wang2015}. 

A major unresolved challenge in realizing this vision is coupling small mode volume mechanical resonators to superconducting quantum circuits. In this work, we analytically and numerically study the coupling between realistic acoustic devices made from piezoelectric thin films to superconducting qubits. Remarkably, we find that strong coupling between a transmon qubit~\cite{Koch2007, Schreier2008} -- a small Josephson junction shunted by a large capacitance -- and a single acousic-wavelength-scale microwave frequency mechanical resonance is achievable. To understand this result, we first discuss a simplified analytical model and derive the scaling of the coupling with resonator size. We then combine microwave network synthesis techniques~\cite{Foster1924a, Nigg2012} with finite-element simulations of piezoelectric nanostructures to fully capture the interaction of quantum circuits with mechanical components. Finally, we design and present full-field piezoelastic simulations~\cite{comsol2013} of Lamb-wave and phononic-crystal resonances and study their scaling properties and the hybridization of their excitations with a transmon qubit.

\begin{figure}[ht]
    \centering
    \includegraphics[width=\linewidth]{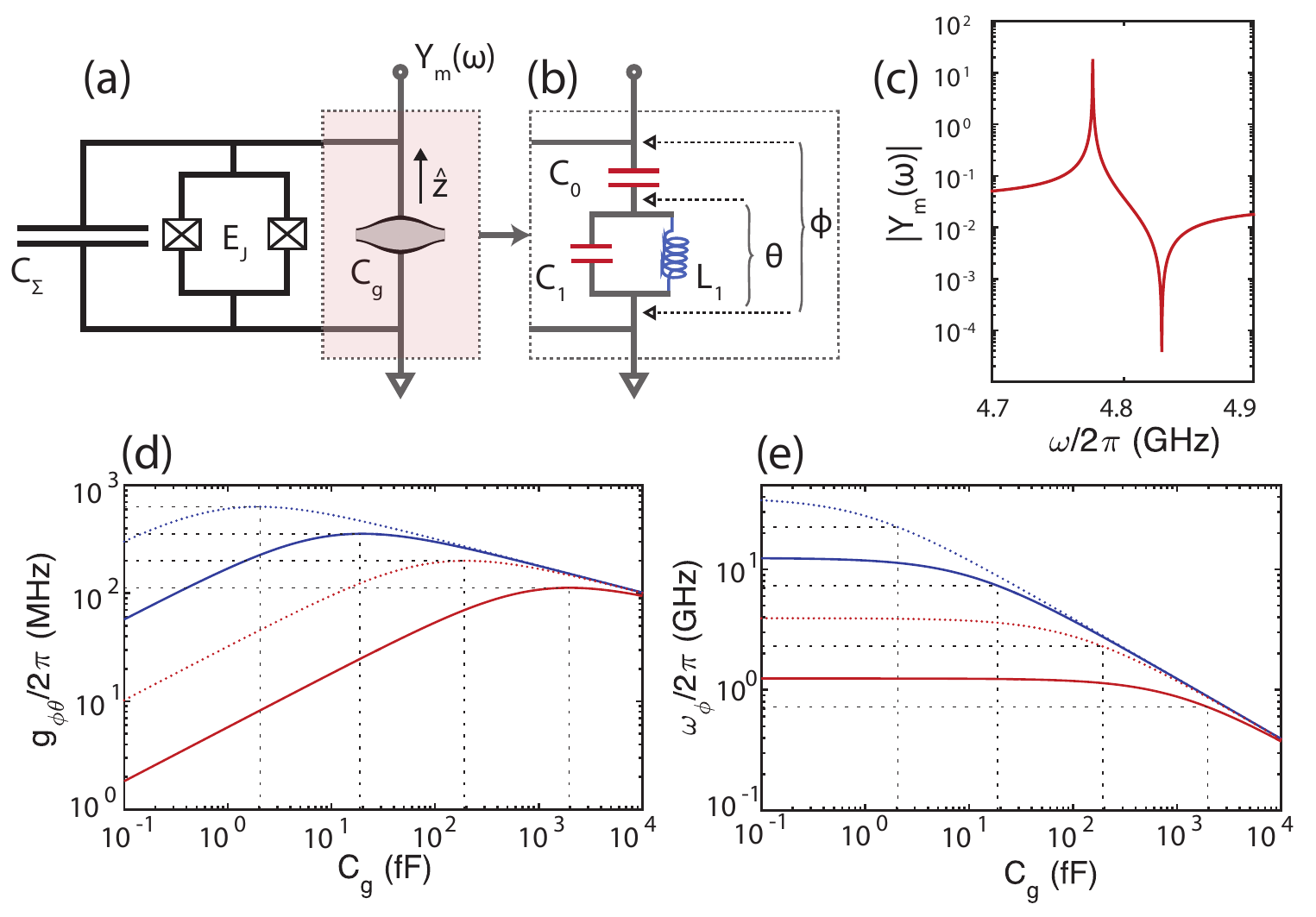}
    \caption{Thin-film bulk acoustic wave resonator. (a) A transmon with Josephson energy $E_J$ and total capacitance $C_\Sigma$ is shunted by a mechanical resonator (highlighted) described by an admittance function $Y_m(\omega)$; (b) Foster network synthesized from the electroacoustic admittance $Y_m(\omega)$ in the vicinity of the fundamental mechanical mode; (c) Exact admittance for a lithium niobate film of thickness $b = 750 \; \text{nm}$; (d) Dependence of the coupling rate $g_{\phi\theta}$ and (e) transmon frequency $\omega_\phi$ on the gate capacitance $C_g$, with $E_J/\hbar = 2\pi \times 10 \; \text{GHz}$ and $C_\Sigma = 10^0, 10^1, 10^2, 10^3 \, \text{fF}$ (dotted blue, solid blue, dotted red, and solid red, respectively). $C_g$ is changed by modifying only the capacitor area, so the mechanical frequency remains unaffected. The maxima of $g_{\phi \theta}$ occur when $C_0 \approx 2C_\Sigma$ (dashed black lines). As the mechanical resonator is shrunk to the regime $C_g \ll C_\Sigma$ the coupling rate falls off sublinearly as $g_{\phi \theta} \sim \sqrt{C_g}$.}
    \label{fig:FBAR}
\end{figure}

\emph{Analytical model}. Consider a piezomechanical resonator coupled to a charge qubit. The system, highlighted in Fig. \ref{fig:FBAR}(a), consists of a thin film of piezoelectric material sandwiched between two electrodes. Ignoring edge effects, there is only one relevant dimension (the $z$ direction normal to the plates) and the system admits an analytic solution \cite{Hashimoto2009, Cleland2004a} for the electrically coupled motional degree of freedom. The current induced on the electrodes is related to the voltage across them through an admittance function
\begin{equation}
\label{FBAR_admittance}
Y_m(\omega) = i\omega C_g\left[1 - K^2 \frac{\tan (\omega b/2\bar{v})}{\omega b/2\bar{v}}\right]^{-1},
\end{equation}
where $K^2 = e_\text{pz}^2/\bar{c}\epsilon$, $e_\text{pz}$ is the piezoelectric coupling coefficient in stress-charge form, $\bar{c} = c + e_\text{pz}^2/\epsilon$ is the modified elasticity coefficient, $b$ is the thickness of the film, and $\bar{v} = \sqrt{\bar{c}/\rho}$ is the speed of sound in the crystal. This exact admittance fully describes the response of the resonator. At frequencies much lower than the fundamental dilational mode of the membrane, \ie time-scales longer than the time it takes sound to travel a distance $b$, the primary contribution to the admittance function is the static electroelastic response, captured by the effective capacitance $C_g/(1-K^2)$. At higher frequencies, elastic waves are excited and the electrical response is modified. This is captured by the second term in Eq. (\ref{FBAR_admittance}), which results in a series of poles and zeros of $Y_m(\omega)$. The first zero at $\omega = 2\pi \bar{v}/2b \equiv \Omega$ corresponds to the fundamental electroacoustic mode of the system, with an associated pole at $\Omega_p < \Omega$. This pole-zero pair is shown in Fig~\ref{fig:FBAR}(c). All subsequent pairs correspond to higher order excitations of the film. 

Knowledge of $Y_m(\omega)$ is sufficient to obtain the full Hamiltonian of the system~\cite{Nigg2012}. Using Foster's theorem \cite{Foster1924a}, we synthesize an electrical linear lossless network that approximates the electroacoustic admittance (see Appendix \ref{app:foster_synth} for details). Restricting our attention to the fundamental mechanical mode, the Foster network becomes the three-node circuit shown in Fig. \ref{fig:FBAR}(b). The zero of the admittance is made explicit for this choice of synthesis, corresponding simply to $\Omega = 1/\sqrt{L_1 C_1}$. Here $C_1  = \lim_{\omega \to \Omega}\{ \frac{1}{2} \text{Im}[\partial_\omega Y_m(\omega)] \} = (C_g/2)(\pi/2K)^2$ is the effective mode capacitance and $C_0 = \lim_{\omega \to 0}\{\text{Im}[\partial_\omega Y_m(\omega)]\} = C_g/(1 - K^2)$ is the electrostatic capacitance of the system, including the elastic contribution. In the limit of vanishing piezoelectric coupling ($K \to 0$), $C_1 \to \infty$ and $C_0 \to C_g$, so the network simply becomes the gate capacitance $C_g$ --- the mechanical resonator becomes invisible to the electrical terminals.

It is now straightforward to derive the Hamiltonian for the coupled transmon-resonator system. Starting from the circuit Lagrangian \cite{Devoret1997, Girvin2011} in terms of the generalized flux variables $\phi$ and $\theta$, defined in Fig. \ref{fig:FBAR}(b), we arrive at the quantized Hamiltonian (see Appendix~\ref{zms_derivation}) 
\begin{multline}
\label{phonon-transmon-h}
\hat{H} = [4E_C^{(\phi)} (\hat{n}_\phi - n_g)^2 - E_J\cos\hat{\phi}] \\ + \hbar\Omega \aopd\aop  + 8E_C^{(\phi,\theta)} n_{\mathrm{zp}}^\theta (\aop + \aopd)(\hat{n}_\phi - n_g).
\end{multline}
Here $\aopd \; (\aop)$ is the creation (annihilation) operator for the phononic mode described by the circuit variable $\theta$ and $\Omega$ is the phonon frequency, $\Ephi$ and $E_J$ are the charging and Josephson energies for the transmon variable $\phi$, $\Ephitheta$ is the cross-charging energy between the $\phi$ and $\theta$ nodes, and $n_\text{zp}^\theta$ is the magnitude of the zero-point charge fluctuations associated with $\theta$. In the transmon limit $E_J/\Ephi \gg 1$, the gate charge $n_g$ can been eliminated by a gauge transformation~\cite{Koch2007}. Further, we can define operators $\bop, \, \bopd$  for the harmonic oscillator approximating the transmon \cite{Koch2007}. The transition frequency for the first two transmon levels is $\omega_\phi = \sqrt{8E_J \Ephi}/\hbar$, and the coupling term becomes
\begin{equation}
\label{coupling_term}
\hat{H}_\text{int} = - i\hbar g_{\phi \theta} (\aop + \aopd)(\bop - \bopd), 
\end{equation}
where
\begin{equation}
g_{\phi \theta} = 8 E_C^{(\phi, \theta)} n_{\mathrm{zp}}^\theta n_{\mathrm{zp}}^\phi/\hbar
\end{equation}
is the rate that sets the phonon-transmon interaction strength. This rate depends only on fundamental constants, the transmon parameters, and the network parameters $(C_0, C_1, L_1)$ which can be readily computed from $Y_m(\omega)$. 

In Figs. \ref{fig:FBAR}(d) and (e), we plot the coupling rate $g_{\phi\theta}$ and the transmon frequency $\omega_\phi$ over many orders of magnitude of the gate capacitance $C_g$ and transmon capacitance $C_\Sigma$. Here we change $C_g$ by changing the area of the gate capacitor while leaving the plate spacing $b$ fixed. This changes the mass of the mechanical oscillator, whereas its frequency remains unmodified. It is interesting to first note that the coupling is maximized at $C_0 = 2C_\Sigma + \mathcal{O}(K^2)$. As we move to larger $C_g$, where $C_g \gg C_\Sigma$, the transducer capacitance dominates so both $g_{\phi \theta}$ and $\omega_\phi$ lose all dependence on $C_\Sigma$ and fall off to zero. Conversely, as $C_g/C_\Sigma \to 0$ the coupling vanishes and $\omega_\phi$ limits to its uncoupled value. Finally, a salient feature of this model is that in the regime $C_g \ll C_\Sigma$, relevant to nanoscale mechanical resonators, the coupling rate scales sublinearly  with the capacitances as $g_{\phi\theta} \sim (C_g^2/C_\Sigma^3)^{1/4}$. This suggests that shrinking down the resonator to a length scale of the order of the acoustic wavelength should be possible without significantly compromising the coupling rate. We remark that Eq. \ref{phonon-transmon-h} is  identical to a circuit QED Hamiltonian \cite{Koch2007} with a transmon qubit, with the cavity photon operator replaced by a phonon operator. 

\begin{figure}
    \centering
    \includegraphics[width=0.475\textwidth]{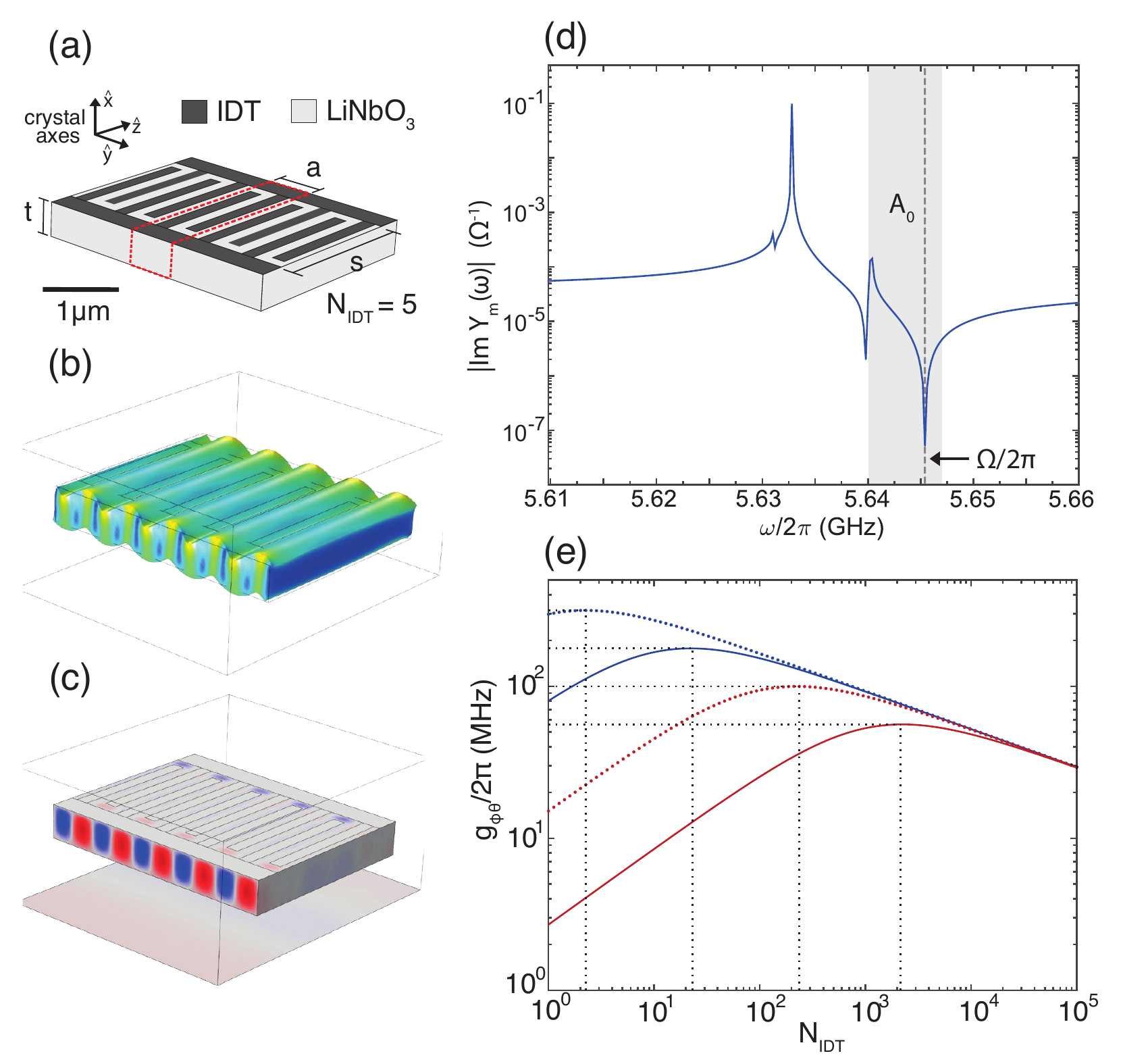}
    \caption{Lamb wave resonator. (a) Simulation geometry for the resonator. An IDT with finger spacing $a = 600 \, \text{nm}$ and $s = 1800 \, \text{nm}$ is patterned over a thin film lithium niobate of thickness $t = 400 \, \text{nm}$. The spacing $a$ can be thought of as a lattice constant (unit cell highlighted). The IDT terminals are driven by an excitation voltage $V(\omega)$ to probe the electroacoustic admittance; (b) Deformation plot for the zeroth-order asymmetric Lamb mode ($A_0$) of wavelength $\lambda = a$; (c) Electrostatic potential distribution generated by the $A_0$ mode. The large overlap between the potential and the electrodes leads to strong piezoelectric coupling to $A_0$; (d) Simulated admittance per unit cell for frequencies near the Lamb mode resonance. The pole and zero corresponding to $A_0$ are shown in the shaded region. The zero located at $\Omega/2\pi = 5.645 \, \text{GHz}$ corresponds to the phononic mode that interacts with the transmon in the single-mode model; (e) Coupling rate to a transmon with $E_J/h = 20 \, \text{GHz}$ as a function of the total number of unit cells $N_\text{IDT}$, with $C_\Sigma = 10^0, 10^1, 10^2, 10^3 \, \text{fF}$ (dotted blue, solid blue, dotted red, and solid red, respectively). }
    \label{fig:lamb_wave}
\end{figure}

We further note that this scaling with capacitor area is fairly general. The interaction energy between the electromagnetic and piezomechanical resonators is given by $Q^\textrm{(piezo)}_\text{zp} V^\textrm{(qubit)}_\text{zp}$, where $Q^\textrm{(piezo)}_\text{zp}$ is the size of the charge fluctuations on the gate capacitance due to the zero-point motion of the mechanical system and $V^\textrm{(qubit)}_\text{zp}$ is the size of the the voltage fluctuations of the qubit in its ground state. The latter only depends on the qubit frequency and capacitance, and for a fixed $E_J$ will scale as $(C_g+C_\Sigma)^{-3/4}$. The charge fluctuations on the gate capacitance scale as $C_g x_\text{zp} \propto \sqrt{C_g}$ since $x_\text{zp}\propto1/\sqrt{m}\propto1/\sqrt{C_g}$. This implies that the coupling scales as  $C_g^{1/2}$ for $C_g \ll C_\Sigma$ and $C_g^{-1/4}$ for $C_g \gg C_\Sigma$, in agreement with our calculations for both the one-dimensional model presented above, and the full-field simulations of a Lamb wave resonator presented below and shown in Figure~\ref{fig:lamb_wave}(e).


\emph{Lamb wave resonator}. Lamb wave modes of films driven by interdigitated (IDT) transducers have found applications in classical information processing and sensing \cite{Zou2014}. In the context of quantum acoustics, since these devices are fabricated on suspended thin films, all phonons are confined in two dimensions and phonon tunnelling into the bulk, which is a loss mechanism in SAW devices~\cite{Gustafsson2014a, Aref2016}, can be eliminated. The system is shown in Fig. \ref{fig:lamb_wave}(a). When the IDT has many periods, there exists a resonance with a frequency that is nearly independent of the number of unit cells, $N_\text{IDT}$. We can therefore explore the dependence of the coupling rate $g_{\phi \theta}$ on the effective gate capacitance $C_g \propto N_\text{IDT}$ for this well-defined class of acoustic modes and benchmark a realistic design as well as its scaling properties.

A finite-element simulation reveals the zeroth-order asymmetric Lamb wave mode ($A_0$) at $5.64 \, \text{GHz}$ (Fig. \ref{fig:lamb_wave}(b)), with a voltage distribution  mode-matched to the IDT fingers (Fig. \ref{fig:lamb_wave}(c)). The mode frequency is set by the finger spacing $a$ and the phase velocity in the $x$ direction, and is weakly dependent on the lateral resonator dimensions. In Fig. \ref{fig:lamb_wave}(d), we show the admittance $Y_m(\omega)$ seen by the circuit terminals at frequencies near the $A_0$ resonance, obtained from finite-element frequency response simulations to compute the current induced in response to an excitation voltage (see Appendix \ref{app:FEM} for details on the simulations). Fitting $Y_m(\omega)$ to a complex rational function \cite{Gustavsen1999a}, we then synthesize the Foster network that reconstructs the Lamb mode admittance, as detailed in Appendix~\ref{app:foster_synth}. Under a simplified model in which only the $A_0$ mode is relevant to the physics, the network is the same as that in Fig. \ref{fig:FBAR}(b). Plotting $g_{\phi \theta}$ as a function of $N_\text{IDT}$ (Fig. \ref{fig:lamb_wave}(e)) then reveals the same scaling derived from the analytical model. In particular, even for a wavelength-scale resonator with $\sim 10$ unit cells, the coupling rate approaches $2\pi \times 50 \, \text{MHz}$ for values of $C_\Sigma$ typically used in transmon qubits.

\begin{figure*}
    \centering
    \includegraphics[width=\textwidth]{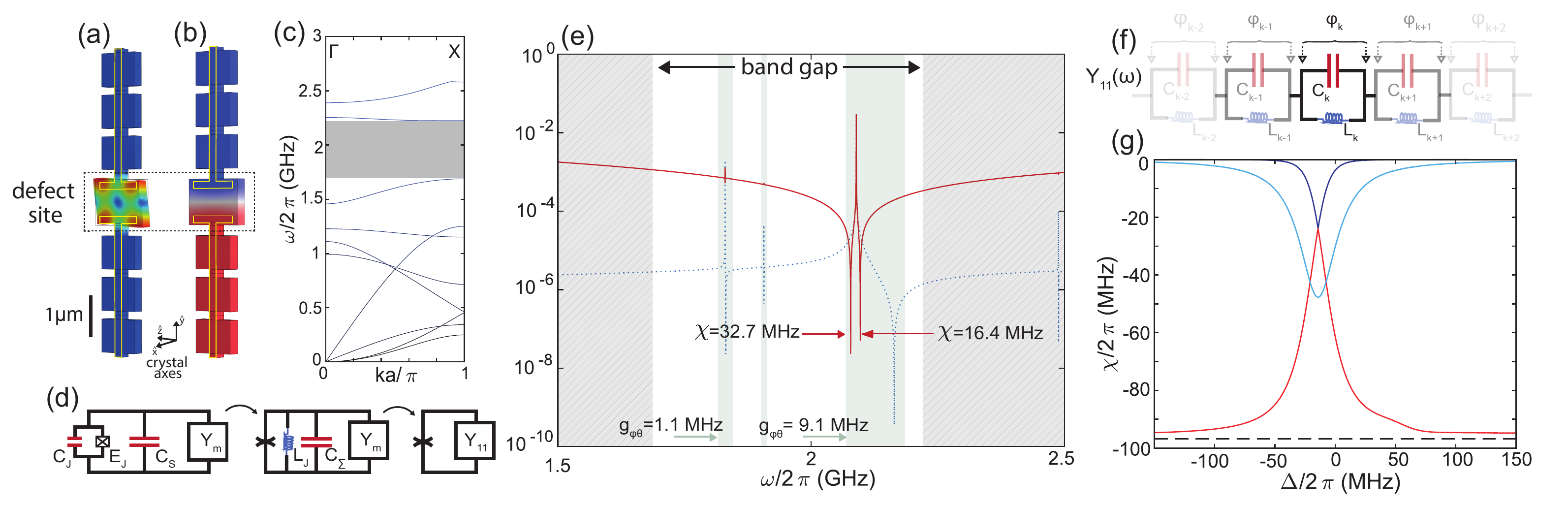}
    \caption{Phononic crystal defect cavity. (a) Deformation plot for the mode of interest at $\Omega/2\pi = 2.089 \, \text{GHz}$, showing a mode tightly localized to the defect region; (b) Electrostatic potential generated by the eigenmode, with a large gradient perpendicular to the symmetry plane at the center of the block (electrical terminals highlighted); (c) Band diagram for the mirror region surrounding the defect site. A full bandgap centered near $2 \, \text{GHz}$ (shaded area) leads to strong confinement of any defect mode lying within this frequency band; (d) In black-box quantization, a transmon with junction parameters $E_J, C_J$ and shunt capacitance $C_S$ is wired to a system described by an admittance function $Y_m(\omega)$. The total capacitance $C_\Sigma = C_J + C_S$ and the linear part of the junction inductance $L_J$ are lumped into the black-box, resulting in a total input admittance $Y_{11}(\omega)$ shunting the nonlinear part of the junction (indicated by the spider symbol); (e) Bare electroacoustic admittance (dashed blue) of the defect site \emph{only}, and total admittance $Y_{11}(\omega)$ (solid red) including the loading from a transmon with $C_\Sigma = 200\,\text{fF}$, $C_J = 2.5\,\text{fF}$, and frequency $\omega/2\pi = 2.1 \, \text{GHz}$. The strongly coupled mode with a pole at $\Omega/2\pi = 2.089 \, \text{GHz}$ can be clearly observed in the admittance spectrum, along with other weakly coupled localized modes; (f) Foster network for $Y_{11}(\omega)$; (g) Anharmonicites of the phonon-like (blue) and transmon-like (red) polaritons. As $\Delta$ is tuned to zero, the phonon-like mode becomes strongly anharmonic and the cross-Kerr term (light blue) is maximized.}
    \label{fig:defect_site}
\end{figure*}

\emph{Phononic crystal defect cavity.} The fact that large coupling rates $g_{\phi\theta}$ are achievable with small gate capacitances opens the design space to study coupling to highly confined acoustic resonances. We consider a localized mode formed by engineering a defect site in a quasi-one-dimensional phononic crystal (Fig. \ref{fig:defect_site}(a) and (b)). Such a phononic crystal can support a large mechanical bandgap~\cite{Safavi-Naeini2010}, as shown in Figure~\ref{fig:defect_site}(c). This bandgap represents a range in frequency where propagation of all elastic waves is disallowed. By creating a defect in this structure, an acoustic mode with frequency inside the bandgap is localized. The characteristic length of this resonator is  $\sim 1~\mu\text{m}$, on the order of the acoustic wavelength. Therefore its mode structure is much simpler and we can consider the full effect of coupling several acoustic modes to the same electrical terminals. 

To perform the multimode analysis we proceed along the lines of black-box quantization \cite{Nigg2012}. This technique --- outlined schematically in Fig. \ref{fig:defect_site}(d) --- consists of lumping the linear part of the transmon into the electroacoustic admittance $Y_m(\omega)$ and synthesizing a Foster network from the total input admittance $Y_{11}({\omega})$ seen by the junction. For the choice of synthesis shown in Fig. \ref{fig:defect_site}(f), each of the $LC$ blocks in the chain corresponds to a normal mode of the transmon-resonator system. This corresponds to diagonalizing the linear part of the total Lagrangian into polariton modes.

The Hamiltonian is
\begin{equation}
\label{multimode_hamiltonian}
\hat{H} = \hbar \sum\limits_k \omega_k \aopd_k \aop_k - E_J\cos\left[\sum\limits_k \psi_{k}^{(\text{zp})} \left(\aop_k + \aopd_k\right) \right],
\end{equation}
where the operators $\{\aop_k^\dagger, \aop_k\}$ create and annihilate the polaritonic excitations of the system \cite{Nigg2012}. In the $E_J/E_C \gg 1$ regime, the zero-point fluctuations $\psi_{k}^{(\text{zp})}$ for the phases are small, and Eq. \ref{multimode_hamiltonian} reduces to an effective Hamiltonian 
\begin{equation}
\label{multimode_hamiltonian_2}
\hat{H} = \hbar \sum\limits_k \omega'_k \aopd_k \aop_k + \frac{1}{2} \hbar \sum\limits_{k j} \chi_{kj} \aopd_k\aop_k\aopd_j\aop_j
\end{equation}
capturing the two-phonon nonlinearities of the modes. Here the $\{\omega'_k\}$ are renormalized frequencies due to a Lamb-shift correction, and the $\{\chi_{kj}\}$ are the anharmonicites which can be computed from the Foster network parameters \cite{Nigg2012}). In Fig. \ref{fig:defect_site}(e), we indicate the coupling rates $g_{\phi \theta}$ for two localized acoustic modes, computed from the single-mode model of Eq. (\ref{phonon-transmon-h}). These show that one mode is indeed much more strongly coupled to the transmon, as already suggested from the admittance spectra. For this mode, we show values of $\chi_{kj}$ for the transmon-like and phonon-like polaritons for a bare transmon frequency $\omega_\phi/2\pi = 2.1 \, \text{GHz}$, slightly detuned from the $\Omega/2\pi = 2.089 \, \text{GHz}$ acoustic mode. We see that the transmon-like polariton remains strongly anharmonic but also contribute a large anharmonicity to the phonon-like polariton. 

We can further study the dependence of the polariton anharmonicites on the detuning $\Delta = \omega - \Omega$, which we sweep by tuning the Josephson energy $E_J$. In Fig. \ref{fig:defect_site}(g), we show the anharmonicities of the transmon-like and phonon-like polaritons as a function of $\Delta$. At large detunings, the phonon mode is essentially linear and the transmon mode has an anharmonicity that asymptotes to its uncoupled value $E_C \simeq e^2/2C_\Sigma$ \cite{Koch2007} (dashed black line in figure). As the two modes become close to resonant, the coupling between the transmon and phonon causes mixing between the modes and the phonon-like mode obtains a large Kerr nonlinearity. Both the value of the self-Kerr ($\chi_{kk}$) and cross-Kerr ($\chi_{jk}$) nonlinearities are plotted in Fig~\ref{fig:defect_site}(g). The maximum value of the anharmonicitiy of the phonon mode is a quarter of the maximum transmon anharmonicity, as expected from a simple hybridization model, and this leads to a $\chi/2\pi \approx 24 \, \text{MHz}$ for the phonon-like mode near $\Delta = 0$.

In conclusion, we have numerically shown that large coupling can be achieved between thin-film, small mode volume mechanical resonators and superconducting microwave quantum circuits despite their vastly different length scales. We have also demonstrated a recipe for calculating these couplings for arbitrary phononic structures. In addition to opening new ways of using acoustic devices in quantum circuits, we expect our results to be directly relevant to transducer designs for piezo-optomechanical devices currently being pursued for microwave to optical conversion~\cite{Bochmann2013a,Balram2015}.

This work was supported by the Stanford Terman Fellowship, ONR MURI QOMAND, a Stanford Graduate Fellowship, as well as start-up funds from Stanford University.



\clearpage

\onecolumngrid

\appendix
\section{Single-mode Hamiltonian} \label{zms_derivation}
Starting with the circuit Lagrangian for a transmon coupled to a single-mode network (see Fig. \ref{fig:FBAR} in the main text),
\begin{equation}
\label{zms_lagrangian}
L = \frac{1}{2}C_0(\dot{\phi} - \dot{\theta})^2 + \frac{1}{2}C_1\dot{\theta}^2 + \frac{1}{2}C_\Sigma \dot{\phi}^2  + E_J \cos \phi - \frac{1}{2L_1}\theta^2,
\end{equation}
the canonical momenta are
\begin{align}
\label{phi_momentum}
\pi_\phi &= \partial_{\dot{\phi}}L = C_0 (\dot{\phi} - \dot{\theta}) + C_\Sigma \dot{\phi} \\
\label{theta_momentum}
\pi_\theta &= \partial_{\dot{\theta}}L = -C_0 (\dot{\phi} - \dot{\theta}) + C_1 \dot{\theta}.
\end{align}
Writing Eqs. (\ref{phi_momentum}) \& (\ref{theta_momentum}) in matrix form, we have
\begin{equation}
\left(
\begin{array}{cc}
C_0 + C_\Sigma  &  - C_0  \\
-C_0       &  C_0 + C_1
\end{array}
\right)
\left(
\begin{array}{cc}
 \dot{\phi}  \\
 \dot{\theta}
 \end{array}
 \right) = 
 \left(
 \begin{array}{cc}
    \pi_\phi  \\
     \pi_\theta
 \end{array}
 \right).
\end{equation}
Inverting the the matrix, we obtain the time derivatives in terms of the momenta
\begin{align}
\label{inv_relations}
\dot{\phi} &= \frac{C_0 + C_1}{C_d^2}\pi_\phi + \frac{C_0}{C_d^2}\pi_\theta \\
\dot{\theta} &= \frac{C_0}{C_d^2}\pi_\phi + \frac{C_0 + C_\Sigma}{C_d^2}\pi_\theta,
\end{align}
where $C_1C_0 + C_\Sigma C_0 + C_\Sigma C_1 \equiv C_d^2$ is the determinant of the matrix. Substituting these relations into Eq. (\ref{zms_lagrangian}), and taking the Legendre transform $H = \pi_\phi \dot{\phi} + \pi_\theta \dot{\theta} - L$, we obtain
\begin{equation}
\label{zms_hamiltonian}
H = \frac{1}{2C_{01}^\Sigma}\pi_\phi^2 + \frac{1}{2C_{0\Sigma}^1}\pi_\theta^2 - \frac{\beta}{C_{1\Sigma}^0}\pi_\theta \pi_\phi - E_J\cos\phi + \frac{1}{2L_1}\theta^2.
\end{equation}
Here we introduced the notation $C_i + (C_j^{-1} + C_k^{-1})^{-1} \equiv C_{jk}^i$ for the equivalent capacitance formed by capacitances $j$ and $k$ in series, in parallel with $i$; $\beta = C_0/C_1^\Sigma$ (in our notation, $C_1^\Sigma = C_1 + C_\Sigma$) is a participation ratio that will determine the phonon-transmon coupling.

To write down the Hamiltonian in the more familiar circuit QED notation, we define the dimensionless charges $n_\phi = \pi_\phi/2e$, $n_\theta = \pi_\theta/2e$, the charging energies $E_C^{(\phi)} = e^2/2C_{01}^\Sigma$, $E_C^{(\theta)} = e^2/2C_{0 \Sigma}^1$, $E_C^{(\phi, \theta)} = \beta e^2/2C_{1\Sigma}^0$, and the inductive energy $E_L = \Phi_0^2/L_1$, where $\Phi_0 = \hbar/2e$ is the reduced flux quantum. Finally, we quantize the degrees of freedom and obtain
\begin{equation}
\hat{H} = 4E_C^{(\phi)} (\hat{n}_\phi - n_g)^2 - E_J \cos\hat{\phi} + 4E_C^{(\theta)} \hat{n}_\theta^2 + \frac{1}{2}E_L\hat{\theta}^2 + 8E_C^{(\phi, \theta)} (\hat{n}_\phi - n_g) \hat{n}_\theta,
\end{equation}
where we have introduced an additional gate charge $n_g$ because, due to the topology of the circuit, the spectrum of $\hat{n}_\phi$ is discrete \cite{Koch2009}. We can write the Hamiltonian in a more familiar form by defining the harmonic oscillator quadratures
\begin{align}
\hat{n}_\theta &= n_{\mathrm{zp}}^\theta (\aop + \aopd) \\
\hat{\theta} &= i\theta_{\mathrm{zp}} (\aopd - \aop), 
\end{align}
with
\begin{align}
n_{\mathrm{zp}}^\theta &= \frac{1}{2}\left(\frac{E_L}{2E_C^{(\theta)}}\right)^{1/4} \\
\theta_{\mathrm{zp}} &= \left(\frac{2E_C^{(\theta)}}{E_L}\right)^{1/4},
\end{align}
so that
\begin{equation}
\label{transmon-phonon-Hamiltonian}
\hat{H} = [4E_C^{(\phi)} (\hat{n}_\phi - n_g)^2 - E_J\cos\hat{\phi}] + \hbar\Omega \aopd\aop  + 8E_C^{(\phi, \theta)} n_{\mathrm{zp}}^\theta (\aop + \aopd)(\hat{n}_\phi - n_g).
\end{equation}
The term in brackets is nothing more than the transmon Hamiltonian, the second term describes a harmonic oscillator with frequency $\Omega = (L_1C_{0\Sigma}^1)^{-\frac{1}{2}} \approx (L_1 C_1)^{-\frac{1}{2}}$, and the third term is a coupling between the oscillator position and the transmon charge. We therefore identify the circuit variable $\phi$ as the transmon degree of freedom, and the $\theta$ variable as the phonon degree of freedom. Going to the transmon limit $E_J/\Ephi \gg 1$, where the zero-point fluctuations of $\phiop$ are small, we can expand the $\cos\phiop$ term in Eq. (\ref{transmon-phonon-Hamiltonian}) to quartic order and define the approximate transmon quadratures \cite{Koch2007, Leib2010}
\begin{align}
\hat{n}_\phi &= i n_{\mathrm{zp}}^\phi (\bopd - \bop) \\
\hat{\phi} &= \phi_{\mathrm{zp}} (\bop + \bopd), 
\end{align}
with
\begin{align}
n_{\mathrm{zp}}^\phi &= \frac{1}{2}\left(\frac{E_J}{2E_C^{(\phi)}}\right)^{1/4} \\
\phi_{\mathrm{zp}} &= \left(\frac{2E_C^{(\phi)}}{E_J}\right)^{1/4},
\end{align}
yielding
\begin{equation}
\hat{H} \simeq [\hbar\omega_\phi \bopd\bop - \frac{E_C^{(\phi)}}{12}(\bop + \bopd)^4] + \hbar\Omega\aopd\aop - i\hbar g_{\phi \theta} (\aop + \aopd)(\bop - \bopd),
\end{equation}
where 
\begin{equation}
\omega_\phi = \sqrt{8E^{(\phi)}_C E_J}/\hbar
\end{equation}
is the transmon frequency and
\begin{equation}
\hbar g_{\phi \theta} = 8 E_C^{(\phi, \theta)} n_{\mathrm{zp}}^\theta n_{\mathrm{zp}}^\phi
\end{equation}
is the coupling energy that sets the phonon-transmon interaction strength.

\section{Finite element simulations} \label{app:FEM}

In order to study resonator designs of arbitrary geometries, we perform full-field finite element method (FEM) simulations to obtain the electroacoustic admittance $Y_m(\omega)$. Using COMSOL Multiphysics~\cite{comsol2013}, we simultaneously solve the equations of elasticity, electrostatics, and their coupling via piezoelectric constitutive relations
\begin{align}
    \label{eq:const-relations}
    D_i &= \epsilon_{i j} E_j + e_{ijk} S_{jk} \\
    T_{ij} &= c_{ijlm} S_{lm} -e_{l ij} E_l,
\end{align}
written in stress-charge form. All repeated indices are summed over. Here $D$ is the electric displacement field, $E$ is the electric field, $T$ is the stress tensor, $S$ is the strain tensor, and $\epsilon$, $c$, and $e$ are the permittivity, elasticity, and piezoelectric coupling tensors, respectively. We can either solve for the eigenmodes of the structure, or perform a frequency response simulation in which an oscillating voltage with amplitude $V(\omega)$ is set on the electrodes as a boundary condition and the field solutions are used to compute the current $I(\omega)$ induced on the electrodes, thereby extracting the admittance 
\begin{equation}
    Y_m(\omega) = \frac{I(\omega)}{V(\omega)}.
\end{equation}
This is conceptually identical to the calculation of $Y_m(\omega)$ in Eq. \ref{FBAR_admittance} in the main text, but is otherwise intractable without numerical tools. \newline

\emph{Lamb wave resonator.} 

The simulation geometry for the Lamb wave resonator consists of a thin layer of X-cut lithium niobate crystal of thickness $t = 400 \; \text{nm}$. The X-cut crystal orientation is implemented by introducing a rotated coordinate system in the simulations (crystal axes labeled in Fig.~\ref{fig:lamb_wave} in the main text). An interdigitated (IDT) capacitor is used to selectively transduce the asymmetric zeroth-order Lamb mode ($A_0$) with a wavelength equal to the IDT finger spacing $a = 600 \; \text{nm}$ (see Fig. \ref{fig:lamb_wave}). In the example discussed in this work, the IDT spans a width $s = 1800 \; \text{nm}$. The terminals are treated solely as voltage boundary conditions on the surface of the LN --- performing more realistic simulations where the terminals are treated as compliant metallic films only slightly modifies the results. Further, the domain of the simulation extends beyond the structure in order to take into account the effect of the fields in vacuum. 

Before calculating $Y_m(\omega)$, we perform eigenmode simulations for rapid characterization of the spectrum and identification of the $A_0$ mode. We also verify the weak dependence of the frequency of $A_0$ on the lateral dimension $s$ and the number of unit cells $N_\text{IDT}$.

In order to simplify the scaling calculations (\emph{i.e.} the calculation of $g_{\phi \theta}$ as a function of $N_\text{IDT}$), we impose Floquet boundary conditions (with $k=0$) on the boundaries of the structure perpendicular to the direction of propagation of the $A_0$ mode. This is equivalent to setting a cyclic boundary condition on a finite one-dimensional crystal --- it eliminates edge effects and guarantees that the frequency of $A_0$ is independent of the total number of unit cells, as independently verified through eigenmode simulations. Due to the periodicity of the structure, the admittance $Y_m^{(N)}(\omega)$ of a resonator with $N$ unit cells is $Y_m^{(N)}(\omega) = NY_m^{(1)}(\omega)$, where $Y_m^{(1)}(\omega)$ is the admittance of a single unit cell. In circuit language, the periodicity allows us to partition the $N_\text{IDT}$ unit cells into $N_\text{IDT}$ parallel networks. This drastically simplifies the calculation, effectively reducing the problem to a numerical simulation of a single unit cell. \newline
 
 \emph{Phononic crystal defect cavity.}
 
\begin{figure*}[ht]
    \centering
    \includegraphics[width=0.8\textwidth]{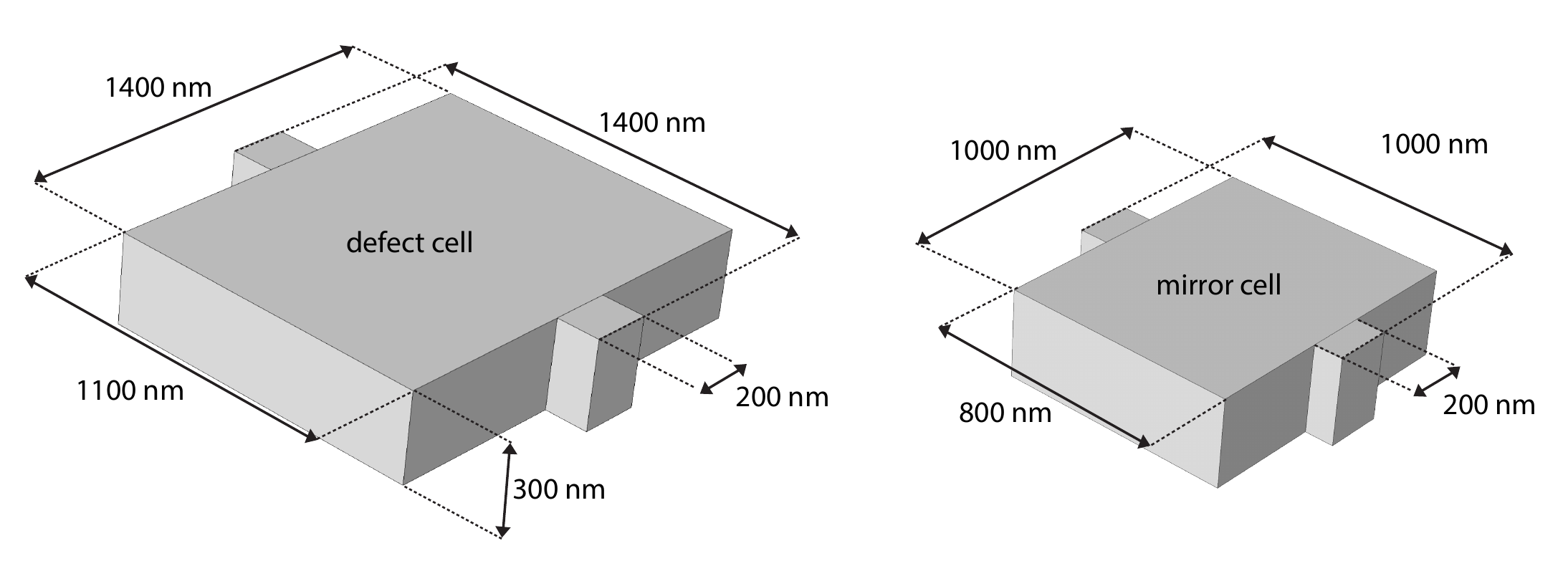}
    \caption{Simulation geomtry for defect site resonator, include both the defect cell and the unit cell for the mirror region.}
    \label{fig:defect_site_SI}
\end{figure*}

The first step in the design flow for the defect cavity is to fully characterize the band structure of the mirror region that supports the bandgap. To this end, we perform eigenmode simulations of a single unit cell with Floquet boundary conditions and sweep the $k$ vector over the one-dimensional Brillouin zone. For a unit cell with dimensions as shown in Fig.~\ref{fig:defect_site_SI}, this generates the band diagram shown in Fig. \ref{fig:defect_site} in the main text. We then design a defect cell with an eigenmode deep inside the bandgap and verify its confinement by simulating the full structure as shown in Figs. \ref{fig:defect_site}(a) and (b). The defect cell dimensions are indicated in Fig.~\ref{fig:defect_site_SI}, and we again use an X-cut crystal orientation (crystal axes labeled in Fig.~\ref{fig:defect_site}). The electrostatic potential generated by the eigenmode has a large gradient perpendicular to the symmetry plane at the center of the block. This motivates placing two electrical terminals that overlap with the blue and red regions in the plot in order to maximize coupling. We then test the mode remains bound after placing voltage terminals that run along the tethers. 

Next we probe the acoustic admittance of the bound mode by simulating the defect site \textit{only}, using fixed boundary conditions at its tethers. For the results shown in Fig. \ref{fig:defect_site}(f) in the main text, we set the resolution $\Delta f$ of the frequency scans to $200 \, \text{kHz}$ in order to fully capture all the eigenmodes within the band of interest. 

\section{Foster synthesis for acoustic systems}\label{app:foster_synth}

A mechanical system with piezoelectric properties, when probed through electrical terminals, is indistinguishable from an ordinary microwave network --- it can be fully described by an admittance function $Y(\omega)$. This has enabled the widespread adoption of mechanical devices as effective circuit elements in classical applications \cite{Campbell1989, Lakin1999, Hashimoto2009}. In this work, we use this insight to abstract away the mechanical aspect of the system and formulate a unified description \emph{at the circuit level}, where calculating coupling rates and other quantities of interest is straightforward. 

\begin{figure*}
    \centering
    \includegraphics[width=0.8\textwidth]{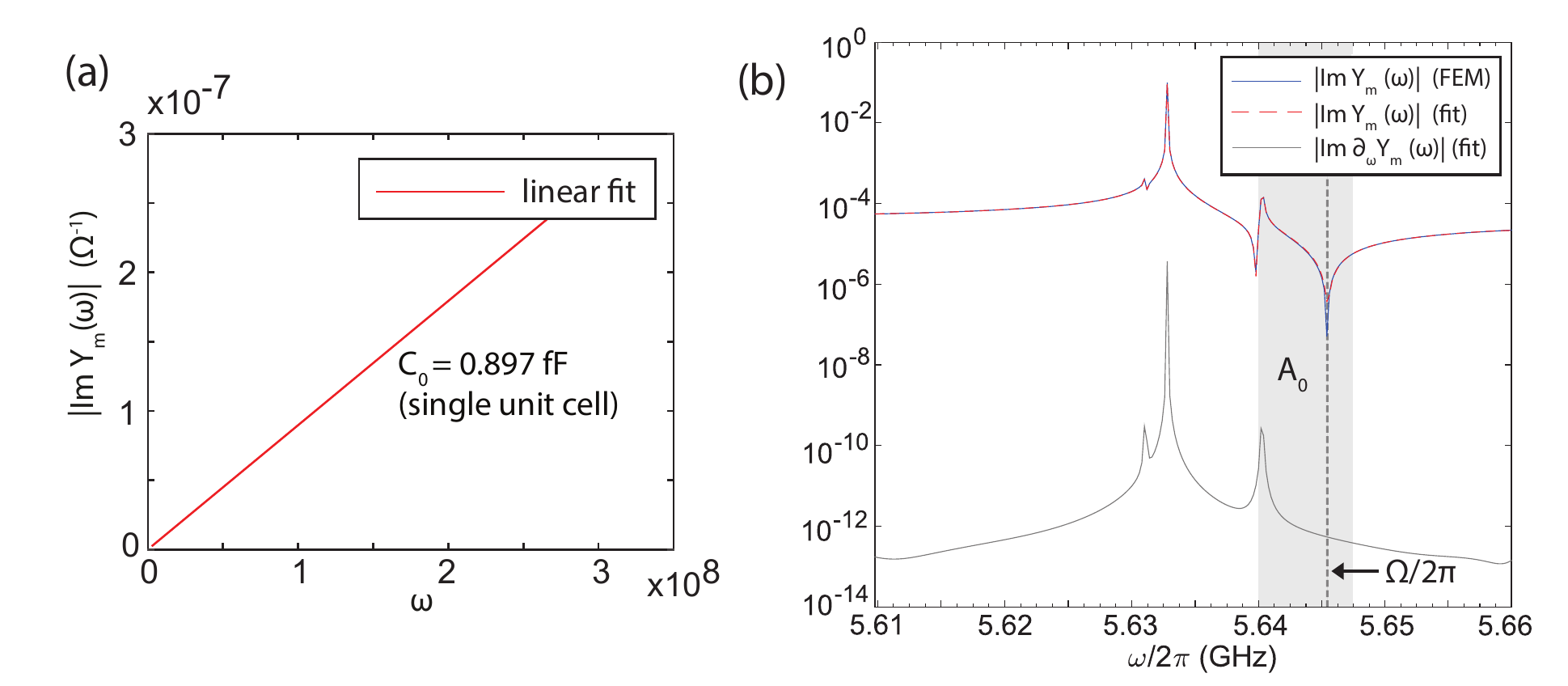}
    \caption{Foster synthesis for Lamb-wave resonator. (a) Electroacoustic admittance near dc, for $\omega/2\pi \in [0, 50]\,\text{MHz}$. The slope of the line is the $C_0$ parameter of the network; (b) Admittance per unit cell near the $A_0$ mode (also shown in Fig.~\ref{fig:lamb_wave}(d) in the main text), including the fit to a rational function and its frequency derivative used to the extract network capacitances.}
    \label{fig:lamb_wave_SI}
\end{figure*}

More precisely, given a linear lossless microwave network with known input admittance $Y_{11}(\omega)$, we would like to explicitly construct a network of capacitances and inductances that is described by the same admittance. Foster synthesis is a well-established technique  \cite{Foster1924a, Guillemin1957} to do this construction. We begin by calculating the acoustic admittance function $Y_m(\omega)$ either analytically or numerically. For a passive, lossless network $Y_m(\omega)$ is a purely imaginary, monotonically increasing function \cite{Foster1924a}. This latter property implies that in general $Y_m(\omega)$ has an alternating sequence of poles and zeros, each corresponding to a resonance and anti-resonance of the network, respectively. A function of this kind can written as a partial fraction expansion of the form 
\begin{equation}
    Y(s) = \sum\limits_k \frac{R_k}{s - s_k} + Cs + D,
\end{equation}
where $s = i\omega + \kappa$ is a complex frequency, $\{s_k\}$ are the poles of $Y(s)$, and $\{R_k\}$ are the associated residues. We fit $Y_m(\omega)$ to a function of this kind using a well-established fitting routine \cite{Gustavsen1999a} (see Fig.~\ref{fig:lamb_wave_SI} for an example of this procedure for the Lamb-wave resonator of Fig.~\ref{fig:lamb_wave}). 

The choice of synthesis is in general not unique. In this work, it is convenient to synthesize $Y_m(\omega)$ as a series combination of parallel inductances and capacitances, such as those shown in Figs. \ref{fig:FBAR} and \ref{fig:defect_site} in the main text. The self-resonance $\omega_k$ of each $LC$ block then corresponds to a zero of $Y_m(\omega)$, and the capacitance $C_k$ can be extracted through
\begin{equation}
C_k = \lim\limits_{\omega \to \omega_k} \left\{ \frac{1}{2}\text{Im}\left[\partial_\omega Y_m(\omega) \right] \right\},
\end{equation}
form which it follows that
\begin{equation}
L_k = \frac{1}{\omega_k^2 C_k}.
\end{equation}

For the single-mode analysis of the analytical model and the Lamb-wave resonator, there is an $LC$ block with $L = \infty$ and $C = C_0$, the dc capacitance of the system (see Fig. \ref{fig:FBAR} in the main text). This is absent in the black-box analysis \cite{Nigg2012} of multimode systems, because there the transmon inductance $L_J$ is lumped into the network as well. To extract $C_0$, we probe the dc response of the system by simulating the admittance spectrum from dc up to $50 \, \text{MHz}$ (Fig.~\ref{fig:lamb_wave_SI}), where $\text{Im}\left[ Y_m(\omega) \right]$ has a featureless linear dependence $\text{Im}\left[ Y_m(\omega) \right] \sim i\omega C_0$ and $C_0$ is simply the slope,
\begin{equation}
C_0 = \lim\limits_{\omega \to 0} \left\{ \text{Im} \left[\partial_\omega Y_m(\omega) \right] \right\}.
\end{equation}

\end{document}